**NANO EXPRESS**  Open Access

# Effect of the anodization voltage on the pore-widening rate of nanoporous anodic alumina

Mohammad Mahbubur Rahman[1], Enric Garcia-Caurel[2*], Abel Santos[1], Lluis F Marsal[1], Josep Pallarès[1] and Josep Ferré-Borrull[1]

**Abstract**

A detailed study of the pore-widening rate of nanoporous anodic alumina layers as a function of the anodization voltage was carried out. The study focuses on samples produced under the same electrolyte and concentration but different anodization voltages within the self-ordering regime. By means of ellipsometry-based optical characterization, it is shown that in the pore-widening process, the porosity increases at a faster rate for lower anodization voltages. This opens the possibility of obtaining three-dimensional nanostructured nanoporous anodic alumina with controlled thickness and refractive index of each layer, and with a refractive index difference of up to 0.24 between layers, for samples produced with oxalic acid electrolytes.

## Background

The research based on nanoporous anodic alumina (NAA) has attracted significant attention in nanoscience and engineering due to its self-assembled, densely packed, and nanoscale-ranged porous structure that naturally forms when aluminum (Al) films are anodized in an acidic electrolyte solution in the appropriate conditions [1,2]. These pores are straight through the film thickness, parallel to each other, and with diameters in the range of 10 to 100 nm. The structural characteristics of the NAA such as pore diameter, interpore distance, porosity, film thickness, and barrier layer thickness are dependent on the anodization conditions [3,4]. These physical properties have made the NAA a suitable material for use as a template to synthesize metal [5,6], polymer [7], and ceramic [8] nanowires and nanotubes. Porous materials, like NAA, have large surface area and specific surface properties that enable them to adsorb watery molecules and thus significantly change their effective refractive index, so chemical and biological sensors have been invented to take advantage of this property [9-12].

Since NAA is a self-assembled porous material with a two-dimensional (2D) pattern having a characteristic interpore distance (which in some fabrication conditions can be in the order of the wavelength of visible light [13,14]), it is possible to control light propagation inside its structure. Its low absorption coefficient, excellent thermal stability, wide electronic band gap (7 to 9.5 eV), and easy handling have made it a potential candidate as a two-dimensional photonic crystal material in the visible and infrared range [15]. Furthermore, nonperiodic nanostructures based on NAA have been demonstrated to show photonic stop bands for all in-plane propagation directions [16,17]. If this in-plane 2D photonic stop band could be combined with vertical optical confinement provided by a periodic change of refractive index in the direction parallel to the pores, then three-dimensional (3D) confinement of light could be achieved. This confinement of light in a small volume and in a porous material could find applications in sensing, LED light extraction, or laser light generation.

Three-dimensional (in-depth) structuring of NAA has drawn great interest for a large range of applications such as high-density storage media [18] or spintronics [19]. Recently, multilayer nanoporous structures or Bragg's stacks have been applied to chemical and

* Correspondence: enric.garcia-caurel@polytechnique.edu
[2]LPICM, Ecole Polytechnique, CNRS, Palaiseau 91128, France
Full list of author information is available at the end of the article biological sensing due to its high reflectivity for a certain incident light wavelength [20-22]. These works report on the fabrication of Bragg mirrors based on NAA with a cyclic porosity with depth, but they do not show control over the optical properties of every cycle. Other authors have also reported on the fabrication of complex pore architectures with modulated pore diameters by cycling between mild and hard anodization conditions [23,24], although the works do not focus on the optical properties of the obtained nanostructures.

In this work, we investigate the possibility of obtaining an in-depth structuring of NAA in layers with controlled thickness and refractive index by using an electrochemical process where the anodization voltage is the only varying parameter, while both the type of electrolyte and its concentration are kept constant. Furthermore, we aim at demonstrating that this can be achieved without changing from mild anodization conditions.

In order to fabricate in-depth structured NAA with significant optical properties, it is necessary to have different layers with the highest possible refractive index contrast. However, it is known that in the self-ordering regime of pore growth, porosity depends weakly on the applied voltage [4,25]. Thus, if a cycling voltage is applied to obtain the NAA, the different layers will have a small refractive index contrast.

Here, we present an innovative approach for obtaining the highest possible refractive index contrast. Our aim is to show that although porosity of as-anodized layers is very similar for all anodization voltages, if a subsequent pore-widening step is applied, the rate at which porosity increases is indeed different. Thus, if a cyclic voltage is applied to obtain NAA, the index contrast between layers obtained with different voltages will be increased with the pore widening.

A simple model can be developed to justify this assumption. The porosity of a NAA layer of vertically straight pores of radius $r$ separated by an interpore distance $D_{int}$ is proportional to the square of the ratio $r/D_{int}$ [4]:

$$p = \alpha \left( \frac{r}{D_{int}} \right)^2 \quad (1)$$

A geometrical analysis reveals that in a perfectly ordered hexagonal structure, the proportionality constant $\alpha$ is $2\pi/\sqrt{3}$. However, if ordering is not perfect, this constant may vary slightly, but in any case, it is weakly dependent on the applied voltage $U$. It is widely accepted that the interpore distance depends linearly on $U$ [4]:

$$D_{int} = kU, \quad (2)$$

where $k$ is the proportionality constant, approximately $k \approx 2.5$ nm/V. Furthermore, the porosity of the as-anodized samples is also weakly dependent on $U$ [4], with a value $P_0$. Thus, from Equations 1 and 2, assuming a perfect triangular ordering, the radius of the as-anodized sample can be written as:

$$r_0 \approx k \sqrt{\frac{\sqrt{3} P_0}{2\pi}} U. \quad (3)$$

The pore-widening process consists of the dissolution of the alumina by 5 wt.% phosphoric acid ($H_3PO_4$). This is a process that takes place at the interface between the alumina and the solvent, and it is reasonable to assume that the pore radius increases linearly with time for constant reaction speed:

$$r(t) = r_0 + \beta t, \quad (4)$$

where the linearity constant $\beta$ depends only on the chemical nature of the alumina and of the solvent and on the process temperature, but there is no reason to think it may depend on the anodization voltage. By substituting this radius in Equation 1, the evolution of the porosity with time can be expressed as:

$$P = \alpha \left( \frac{r_0 + \beta t}{D_{int}} \right)^2. \quad (5)$$

The derivative of this porosity with respect to time corresponds to its rate of growth if we consider small reaction times ($t \to 0$):

$$\left. \frac{dP}{dt} \right|_{t=0} = \frac{2\alpha r_0 \beta}{D_{int}^2} = \frac{2\alpha \beta}{k} \sqrt{\frac{\sqrt{3}}{2\pi} P_0} \frac{1}{U}, \quad (6)$$

where the dependence on the anodization voltage has been made explicit. Thus, the rate of pore widening should be inversely proportional to the anodization voltage.

In the following section, we describe an experiment in order to check this hypothesis, with the details of the fabrication procedure and the characterization methods. Then, the results of the experiment are presented and discussed, and finally, the conclusions are summarized.

## Methods

In order to show that the pore-widening rate is different for NAA obtained with different anodization voltages, an experiment was designed where a set of NAA films are obtained under different applied voltages and pore-widening times, but under the same acid electrolyte and concentration conditions. The NAA films on the Al substrate were fabricated using the well-established two-step anodization method, whose details are described elsewhere [26,27]. In our case, we used high-purity Al substrates (99.99%) of 500 μm thickness from Sigma-Aldrich (St.

Louis, MO, USA). In order to improve the physical properties of the NAA film, the commercial Al substrate needed pretreatment: First, the Al substrates were cleaned by water-ethanol-water and then dried. Then, the surface roughness was reduced by an electropolishing process performed at room temperature and 20 V for 4 min in a 1:4 $v/v$ mixture of perchloric acid and ethanol with continuous stirring. During the electropolishing process, the sense of the stirrer was switched every 1 min. After electropolishing, the samples were cleaned in deionized water. After the pretreatment, the two-step anodization method was performed on the Al surface using 0.3 M oxalic acid ($H_2C_2O_4$) solution at a temperature between 5 and 7°C. The first step of the anodization process was carried out at a constant voltage (V) of 40 V for 20 h. The resulting nanostructure after the first step is a thin film of alumina with disordered pores at the top but self-ordered pores at the bottom [28]. This alumina film was dissolved by wet chemical etching at 70°C in a solution of chromic and phosphoric acid (0.4 M $H_3PO_4$ and 0.2 M $H_3CrO_3$) and stirred at 300 rpm for 3 h. After removing the alumina film, the second anodization step was performed using the same acid electrolyte (0.3 M $H_2C_2O_4$) and the same temperature. We divided our samples into four groups with four samples in each group (Additional file 1: Table S1). Each group of samples was fabricated using a specific applied voltage: 20, 30, 40, and 50 V. These voltages were chosen because this is the range in which self-ordering is obtained for anodization in 0.3 M oxalic acid.

Each sample within a group (all obtained under the same applied voltage) was fabricated by applying a different anodization time: 10, 15, 20, and 25 min (Additional file 1: Table S1), and the charge going through the circuit was recorded by the anodization system control software. Applying different anodization times would permit calibrating the thickness of the NAA film as a function of the total charge used in the anodization procedure.

Finally, pore widening with 5 wt.%. phosphoric acid ($H_3PO_4$) at 35°C was applied to the samples. The pore-widening time varied within a group, being 0, 3, 6, and 9 min for the samples produced with 10-, 15-, 20-, and 25-min anodization, respectively. With this, the samples were obtained for all the range of applied voltages and pore-widening times, and were used to check if the pore-widening rate depends on the anodization voltage.

The NAA films were characterized by environmental scanning electron microscopy (ESEM; FEI Quanta 600, FEI Co., Hillsboro, OR, USA) and ellipsometry (MM16 Polarimeter, HORIBA Jobin-Yvon). The ellipsometry technique permits obtaining the spectra of light polarization change (expressed as the Δ and Ψ angles) upon reflection on the sample at different angles of incidence [29]. Each sample was measured in the wavelength range of 400 to 1,000 nm and for angles of incidence at 60°, 70°, 75°, and 80°. All the spectra for the same sample were analyzed simultaneously by fitting them into theoretical ellipsometry curves obtained from an optical model of the sample, using the DeltaPsi2 software from HORIBA Jobin-Yvon. The optical model of the sample depends on parameters such as the thickness of the NAA film and its porosity. Thus, the best fit can be considered as an estimation of these parameters on the sample. By using ellipsometry, it is possible to obtain an accurate estimation of thin film thickness of about 1 nm up to several microns thick.

The optical model considered for the samples in this work is depicted in Figure 1. It consists of two thin film layers on an aluminum substrate. The top layer is a model for the NAA film with thickness ($d$) and optical properties obtained from the Bruggeman [30] effective medium approximation of a mixture of aluminum oxide and air. This approximation gives the refractive index of an effective medium from the refractive indices of the constituent materials and their corresponding volume fractions. In this case, the volume fraction of air ($P$) corresponds to the porosity of the NAA film. In between this top layer and the substrate, the model is considered to have a very thin interfacial layer composed of a mixture of aluminum, aluminum oxide, and air in order to take into account the nanopatterning of the aluminum substrate at the bottom of the NAA pores and the barrier layer. Including this interfacial layer permitted achieving better fits of the ellipsometric data.

As an example of the optical characterization, in Figure 2, the ellipsometric spectra corresponding to the sample obtained at an anodization voltage of 20 V during 10 min and with no pore widening are shown for the four angles of incidence. The figures also include the best fit obtained from the optical model of the sample. It has to be noted that this best fit is obtained by taking into consideration all the angles of incidence simultaneously. For this sample, the best fit corresponds to a thickness of $d = 202$ nm and a porosity $P = 28\%$. Furthermore, in order to validate the accuracy of the ellipsometry results, cross-sectional ESEM images were used to estimate the thickness of some of the NAA films.

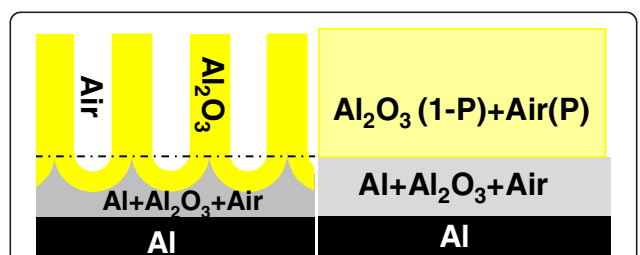

**Figure 1 Sketch of the structure used to model the ellipsometric data.** Left: schematic drawing of the NAA structure showing the pores, the alumina, and the aluminum substrate. Right: the corresponding layered optical model considered by the characterization software. *P* is the porosity of the NAA, which corresponds to the volume fraction of air. The plus sign represents the use of a Bruggeman effective medium approximation to model the refractive index of the mixture of materials in the layer.

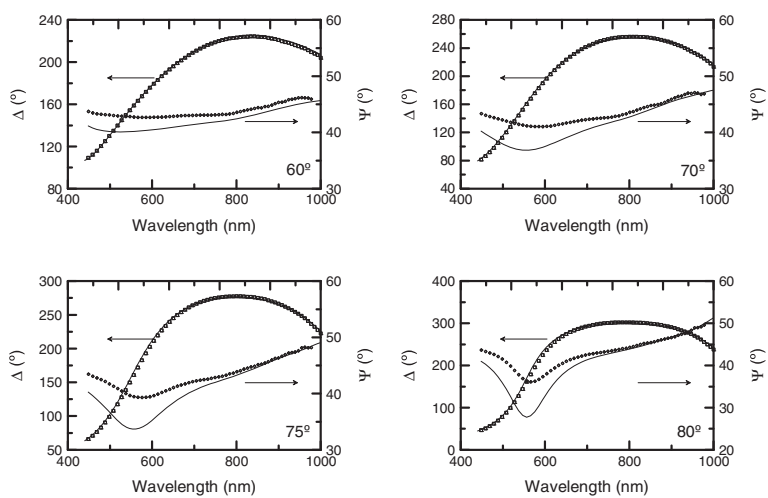

**Figure 2 Ellipsometric measurements for the sample 110928-Al1 (symbols).** Together with the best fit (considering all the angles of incidence simultaneously) obtained with the DeltaPsi2 software. The corresponding measurement angle of incidence is indicated in each graph.

## Results and discussion

Figure 3a,b,c,d shows the cross-sectional and surface ESEM pictures of NAA films anodized at four different voltages: 20, 30, 40, and 50 V, respectively. For all anodization voltages, perfect, hexagonally ordered pores with uniform size on the surface were observed as expected because all the samples were fabricated with the same first-step anodization voltage. The cross sections show uniform straight pores for the samples produced with 40

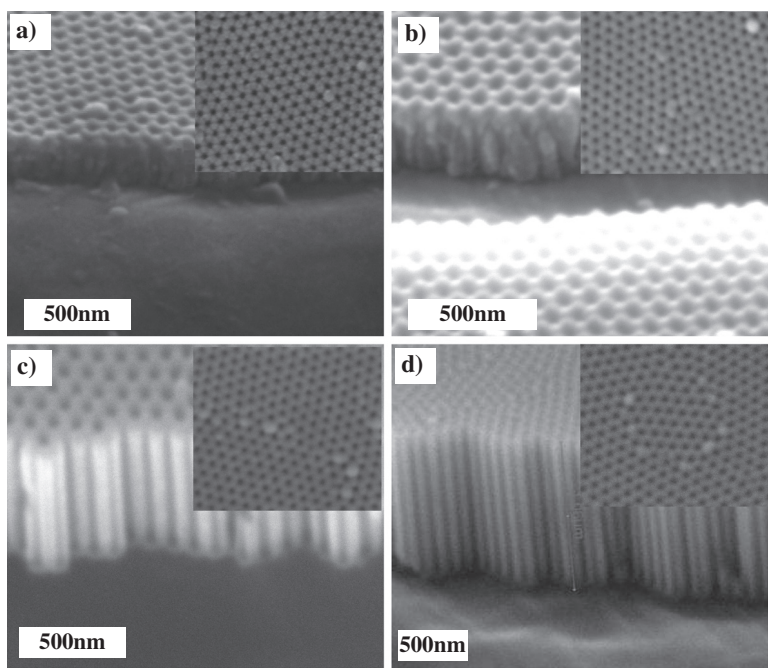

**Figure 3 Set of ESEM images of the as-anodized NAA fabricated with different applied voltages.** (**a**) 20, (**b**) 30, (**c**) 40, and (**d**) 50 V. All the samples were obtained after 10 min of anodization.

and 50 V in the second step. For the samples produced with smaller voltages, the pore uniformity is not as evident as the thickness of the NAA layer is smaller. In order to confirm that the pores reached a steady growth, we checked that the current transient for all the samples reaches a constant value of anodization current (Additional file 1: Figure S1).

The results of the optical characterization are summarized in Figure 4. The resulting best-fitted values of thickness and porosity of each sample are listed in Additional file 1: Table S2. In order to validate the obtained results, the physical thickness for some of the samples was estimated from the ESEM cross-sectional pictures, shown in Additional file 1: Figure S2. A good agreement between the optical characterization and ESEM estimations was achieved.

To show that the pore-widening rate depends on the anodization voltage, we plot the porosities of the studied NAA layers as a function of the pore-widening time (Figure 4a, the dashed lines are a guide to the eye). It can be seen that the porosity of the NAA layer increases at a different rate depending on the anodization voltage, with a bigger rate for the smaller voltage. This is in agreement with our hypothesis that the pore radius increases linearly with pore-widening time for small times, and the porosity growth rate is inversely proportional to the anodization voltage. Furthermore, the results show that the porosity of the as-produced samples is similar and varies between 25% and 28%. This result is in good agreement with the work of Gâlc et al. [31], where similar porosities were found for samples fabricated with the same process. This also confirms the weak dependence of porosity on the anodization voltage as long as the samples are produced in the self-ordering regime [4]. This assumption is also supported by the fact that the NAA layer thickness depends linearly on the total charge spent in anodization regardless of the applied anodization voltage (as depicted in Figure 4b). Such a good linearity between thickness and total charge could be used to control with precision the thickness of layers produced with different voltages in the same electrochemical process to obtain a 3D nanostructure.

Since the goal of this work is to show the possibility of obtaining NAA composed of different layers with different refractive indices, we plot the refractive index of the obtained NAA at a wavelength of 750 nm as a function of the pore-widening time (shown in Figure 4c). The refractive index values were obtained from the optical characterization results applying the Bruggeman effective medium approximation. With the applied voltages and pore-widening times, a refractive index difference of up to 0.24 can be achieved.

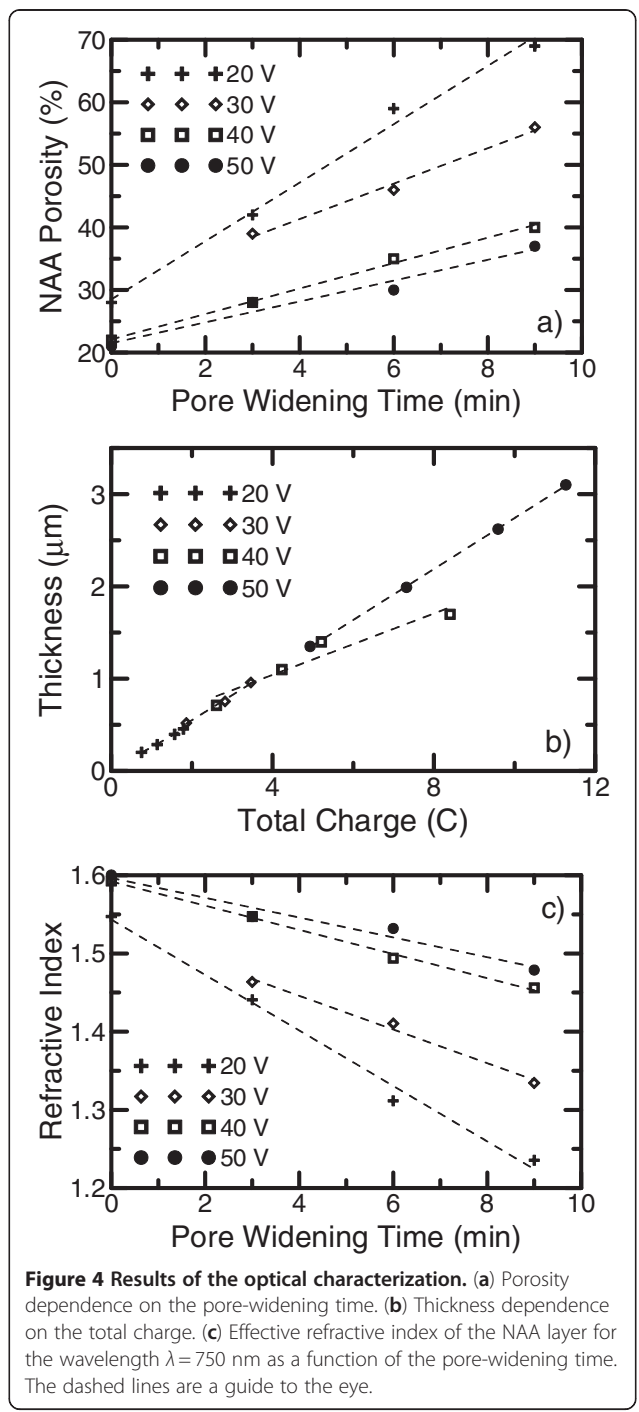

Figure 4 Results of the optical characterization. (a) Porosity dependence on the pore-widening time. (b) Thickness dependence on the total charge. (c) Effective refractive index of the NAA layer for the wavelength $\lambda = 750$ nm as a function of the pore-widening time. The dashed lines are a guide to the eye.

## Conclusions

By means of an ellipsometry-based optical characterization combined with ESEM picture analysis, we have shown that the pore-widening rate of NAA is higher when obtained with a smaller anodization voltage. This has been demonstrated for samples fabricated with the anodization voltage as the only difference but with the same acid electrolyte and concentration conditions. This opens the possibility of combining different voltages in the same electrochemical process to obtain multilayered

NAA nanostructures with a difference in refractive index of up to 0.24 between the layers. Furthermore, this shows that multilayered NAA with remarkable optical properties can be fabricated without switching between mild and hard anodization conditions. These 3D nanostructured NAA can have photonic properties such as the existence of 3D photonic stop bands or the confinement of light in 3D nanocavities.

## Additional file

**Additional file 1: Additional documentation: Contains Tables S1 and S2, and Figures S1 and S2.**

### Competing interests
The authors declare that they have no competing interests.

### Authors' contributions
MMR designed the experiment, fabricated the NAAs, performed the optical characterization, and redacted and revised the manuscript. EGC performed the optical characterization, and manuscript revision. AS fabricated the NAAs and revised the manuscript. LFM and JP revised the manuscript. JFB designed the experiment, performed the optical characterization, and redacted and revised the manuscript. All authors read and approved the final manuscript.

### Authors' information
JFB received his BS, MS, and PhD degrees in physics from the University of Barcelona (Barcelona, Spain) in 1993, 1995, and 1999, respectively. After obtaining his PhD, he spent 1 year and a half at the Fraunhofer Institute (Jena, Germany) and 7 months at the ENEA (Rome, Italy). At present, he is an associate professor at the Department of Electronic, Electrical and Automatic Control Engineering, Universitat Rovira i Virgili (Tarragona, Spain), where he develops his research activities as well as his teaching duties. LFM received his BS and MS degrees from the University of Barcelona (Barcelona, Spain) in 1991 and his PhD degree from the University Politècnica de Catalunya (Barcelona, Spain) in 1997, all in physics. In 1991, he joined the Department of Electronic Engineering, University Rovira i Virgili (Tarragona, Spain), where he became assistant professor in 1994, associate professor in 2000, and full professor in 2009. Nowadays, he is head of the Department of Electronic, Electrical and Automatic Control Engineering. From April 1998 to May 1999, he was a postdoctoral researcher at the Department of Electrical and Computer Engineering, University of Waterloo (Ontario, Canada). JP received his BS and MS degrees in physics from the Universitat de Barcelona (Barcelona, Spain) in 1989 and 1991, respectively, and his PhD degree in physics from the Universitat Politecnica de Catalunya (Barcelona, Spain) in 1997. After obtaining his PhD, he spent 1 year at the Debye Institute, University of Utrecht (Utrecht, The Netherlands). At present, he is a full professor at the Department of Electronic, Electrical and Automatic Control Engineering, Universitat Rovira i Virgili (Tarragona, Spain), where he develops his research activities as well as his teaching duties. EGC finished his PhD in physics from the University of Barcelona, Spain, in 2001. He joined as a permanent researcher at Ecole Polytechnique, France, in 2004. He is involved in the research in ellipsometry and polarimetry for materials characterization and also in the research and development of new optical instruments in collaboration with HORIBA Jobin-Yvon to explore the new applications of polarized light. He is also working as an associate professor in Electronics & Solid State Physics in SUPELEC and also in the Department of Optics in the Telecom & Management Sud-Paris. AS received his chemical engineering degree at the University Jaume I of Castelló de la Plana (Castelló, Spain) in 2005 and his MS and PhD degree in electronic engineering with a scholarship at the Department of Electronic, Electrical and Automatic Control Engineering at the University Rovira i Virgili (Tarragona, Spain) in 2007 and 2010, respectively. MMR completed his BS degree in physics at the Jahangirnagar University Savar (Dhaka, Bangladesh), MS degree in nanoscience at the Chalmers University of Technology (Gothenburg, Sweden), and MS in electronic engineering at the University Rovira i Virgili (Tarragona, Spain). At present, he is carrying out his PhD thesis with a scholarship at the Department of Electronic, Electrical and Automatic Control Engineering, Universitat Rovira i Virgili. The main objective of this work is to develop a technology to obtain photonic quasi-random nanostructures based on nanoporous anodic alumina.

### Acknowledgments
This work was supported by the Spanish Ministry of Science under the projects TEC2009-09551, HOPE CSD2007-00007 (Consolider-Ingenio 2010), and FR2009-0005.

### Author details
[1]Nano-electronic and Photonic Systems (NePhoS), Universitat Rovira i Virgili, Avinguda Paisos Catalans 26, Tarragona 43007, Spain. [2]LPICM, Ecole Polytechnique, CNRS, Palaiseau 91128, France.